\documentclass[a4paper,11pt]{articleSBPO}
\usepackage{sobrapo-template}
\usepackage[utf8]{inputenc}
\usepackage{amsmath,amssymb}
\usepackage[a4paper,top=3.3cm,left=2.9cm,right=2.9cm,bottom=2.5cm,noheadfoot]{}

\usepackage[algoruled,lined,linesnumbered]{algorithm2e}
\usepackage{lscape}
\usepackage{longtable}
\usepackage{indentfirst}
\usepackage{array}
\usepackage{multirow}
\usepackage{enumerate}
\usepackage[figuresright]{rotating}
\usepackage{natbib} 

\title{Solving the Quadratic Assignment Problem on heterogeneous environment (CPUs and GPUs)   with the application of Level 2 Reformulation and Linearization Technique} 

\begin{document}

\maketitle

\begin{center}
\textbf{Alexandre Domingues Gonçalves} \\
Campus São Gonçalo - Instituto Federal do Rio de Janeiro - IFRJ\\
Rua José Augusto Pereira dos Santos s/n - Neves - São Gonçalo - RJ\\
{alexandre.domingues@ifrj.edu.br}\\
\par\end{center}

\begin{center}
\textbf{Artur Alves Pessoa}\\
{Engenharia de Produção - Universidade Federal Fluminense - UFF}\\
{Rua Passo da Pátria 156 - Bloco E - 4º andar - São Domingos - Niterói - RJ}\\
{artur@producao.uff.br}\\
\par\end{center}

\begin{center}
\textbf{Lúcia Maria de Assumpção Drummond}\\
Instituto de Computação - Universidade Federal Fluminense - UFF\\
Rua Passo da Pátria 156 - Instituto de computação - sala 525 - São Domingos - Niterói - RJ\\
{lucia@ic.uff.br}\\
\par\end{center}

\begin{center}
\textbf{Cristiana Bentes}\\
Departamento de Geomática - Universidade Estadual do Rio de Janeiro - UERJ\\
Rua São Francisco Xavier 524 - 5o Andar - Bloco D - Maracanã - Rio de Janeiro - RJ\\
{cris@eng.uerj.com}
\par\end{center}

\begin{center}
\textbf{Ricardo Farias}\\
{COPPE Sistemas - Universidade Federal do Rio de Janeiro - UFRJ}\\
{Centro de Tecnologia - Bloco H - Sala 319 - Ilha do Fundão - Rio de Janeiro - RJ}\\
{rfarias@cos.ufrj.br}\\
\par\end{center}

\vspace{8mm}
\begin{abstract}
The Quadratic Assignment Problem, QAP, is a classic combinatorial optimization problem,  classified as NP-hard and widely studied. This problem consists in assigning N facilities to N locations obeying the relation of 1 to 1, aiming to minimize costs of the  displacement between the facilities. The application of Reformulation and Linearization Technique, RLT, to the QAP leads to a tight linear relaxation but large and difficult to solve. Previous works based on level 3 RLT needed about 700GB of working memory to process one large instances (N = 30 facilities). We present a modified version of the algorithm proposed by Adams et al. which executes on heterogeneous systems (CPUs and GPUs), based on level 2 RLT. For some instances, our algorithm is up to 140 times faster and occupy 97\% less memory than the level 3 RLT version. The proposed algorithm was able to solve by first time two instances: tai35b and tai40b.
\end{abstract}

\bigskip
\begin{keywords}
QAP. RLT. GPU.

\bigskip
\noindent{Main Area: Combinatorial Optimization}
\end{keywords}

\section{Introduction}

The Quadratic Assignment Problem, or QAP, is one of the hardest and studied combinatorial optimization problems of literature. Consists to find an allocation of $N$ facilities to $N$ locations obeying the ratio of 1 to 1 and aiming to minimize the  cost obtained by the sum of flow-distance products.
Its formulation was initially presented by \cite{koopman1957} and has practical applications in object allocation into departments, electronic circuit boards design, layouts problems, construction planning, and others.
For a detailed study on the QAP we suggest the following references: \cite{Pardalos94}, \cite{Padberg96}, \cite{Burkard98} and \cite{cela1997}.

Exact methods to solve the QAP require high computational power, for example, in \cite{hahn2013},  one instance with 30 facilities (as nug30) spent 8 days in a supercomputer composed by 32 nodes of 2.2GHz Intel Xeon CPU and 700GB of main memory. The best known strategy to achieve the exact solution of QAP is using branch-and-bound algorithms where the main problem is divided  into smaller sub-problems in trying to solve them exactly. 

The lower bound are essential components for branch-and-bound procedures because they allow discards a large number of alternatives in the search for the optimal solution. The limits reached by \cite{burer2006}, \cite{adams2007} and \cite{hahn2012} stand out as the most tighter to the QAP. The proposed by \cite{burer2006} consists of relaxations lift-and-project for binary integer problems, \cite{adams2007} presents a dual ascent algorithm based on level 2 reformulation and linearization technique  and \cite{HahnRLT3} that also present a dual ascent algorithm but based on level 3 reformulation and linearization technique.

The reformulation and linearization technique, or RLT, was initially developed by \cite{adams1990}, \cite{sherali1999}, in order to generate linear relaxations with tight lower bounds for a class of mixed integer programming 0-1 problems. In the literature, we find applied to the QAP the level 1 reformulation and linearization technique, or RLT1, in \cite{hahn1998}, the level 2 reformulation and linearization technique, or RLT2, in \cite{adams2007} and  the level 3 reformulation and linearization technique, or RLT3 in \cite{hahn2012}. Parallel versions of RLT3 are proposed in \cite{hahn2013} and \cite{rlt3distribuido} for a cluster with shared memory and a distributed environment, respectively. The mentioned studies have shown that higher the level of RLT used, tighter lower bound is obtained, however, the working memory (RAM) required to execute the algorithm   increases exponentially as grows the level of RLT.

We didn't find studies in the literature addressing exact methods for the QAP solution using a computational structure composed of CPUs and GPUs. We found  heuristics, such as the work of \cite{Tsutsui2011} based on ant colony optimization algorithm combined with Tabu search and another using genetic algorithms in \cite{Tsutsui2009}.

Our proposal consists of an branch-and-bound algorithm that  uses to calculate the lower bound  a modified version of the dual ascent algorithm of \cite{adams2007}  to be executed on a heterogeneous environment (CPUs and GPUs). The algorithm takes advantage of the good bounds reached by RLT2 relaxation and the computational power of GPUs with low use of working memory when compared to RLT3 dual ascent algorithm. The RLT3 algorithm reaches tighter bounds  than the previous but it requires a prohibitive memory for commercial computers.

This work is organized as follow: In the section \ref{sec:qap} we show the QAP formulation. In the Section \ref{sec:rlt} we describe the RLT method and its application to QAP. In the Section \ref{sec:dualascent} we present the dual ascent procedure based in RLT2 and its operations. In the Section \ref{sec:GPU} we detail the implementation of branch-and-bound to be executed on a heterogeneous environment (CPUs and GPUs). In the Section \ref{sec:Resultados} we report the experimental results, and finily, the consideration and future works in Section  \ref{sec:conclusao}.

\section{The Quadratic Assignment Problem (QAP) formulation} \label{sec:qap}
 
Given $N$ facilities, $N$ locations, a flow $f_{ik}$ from each facility $i$ to each facility $k$, $k \neq i$, and a distance $d_{jn}$ from each location  $j$ to each location $n$, $n \neq j$, the QAP consists to assign each facility $i$ to exactly distinct location $j$ in order to find:

\begin{equation}
\label{eq:qap1}  min  \displaystyle  \sum_{i=1}^{N} \sum_{j=1}^{N} \sum_{k=1 \atop {k\not=i} }^{N} \sum_{n=1 \atop {n\not=j}}^{N} f_{ik} d_{jn} x_{ij} x_{kn} \ \
\end{equation}

\begin{equation}
\mbox{s.t.} \label{eq:qap2} \displaystyle \ \sum_{i=1}^{N} x_{ij}  = 1 \ \  \forall  \ \ \  j = 1,...,N \ 
\end{equation}

\begin{equation}
\label{eq:qap3} \ \ \ \ \  \displaystyle \ \sum_{j=1}^{N} x_{ij}  = 1  \ \ \forall \ \  i = 1,...,N \ 
\end{equation}

\begin{equation}
\label{eq:qap4}  \ \ \ \ \  \displaystyle \  x_{ij} \in  \left\{0,1\right\}   \ \ \forall \ \ i = 1,...,N; \ \ \ j = 1,...,N  \\
\end{equation}

\section{The Reformulation and linearization technique - RLT } \label{sec:rlt}

For problems involving $n$ variables, the technique of reformulation and linearization (RLT) provides $n$ hierarchical levels of relaxation for a convex polygon of integer solutions to the original problem. For a given level $k$,  $k \in \{ 1, ..., n \} $, also called as RLT $k$, the technique uses every polynomial factors of degree $k$ involving $k$ binary variables $x$ or its complementaries $(1 – x)$.

The linearization consists in adding auxiliary variables, each representing a possible product of original or complementary variables, and relating them to the original by new restrictions. Each new restriction corresponds to a unique restriction multiplied by a product of unique or additional variables.
Assuming  $c_{ijkn} = f_{ik}d_{jn}$ and $d_{ijknpq} = 0$ for all $i,k,p=1 ,...,N$ with distinct $i,k,p$, and $j,n,q = 1,...,N$, also with distinct $j,n,q$. The following formulation is obtained when applying the method RLT2 to the QAP formulation (equations (\ref{eq:qap1}-\ref{eq:qap4})), as \cite{adams2007};

\begin{equation}
\label{eq:rlt2a}  min \left\{ \displaystyle \sum_{i=1}^{N} \sum_{j=1}^{N} b_{ij}x_{ij} +  \sum_{i=1}^{N} \sum_{j=1}^{N} \sum_{k=1 \atop{ k\not= i}}^{N} \sum_{n=1 \atop{n\not= j}}^{N} c_{ijkn} x'_{ijkn} + 
   \displaystyle  \sum_{i=1}^{N} \sum_{j=1}^{N} \sum_{k=1 \atop { k\not= i}}^{N} \sum_{n=1 \atop{n\not= j}}^{N} \sum_{p=1 \atop{ p\not= i,k}}^{N} \sum_{q=1 \atop{q\not= j,n}}^{N}d_{ijknpq} x''_{ijknpq}    \right\}
\end{equation}

\noindent
\begin{equation}
\mbox{s.t.}\label{eq:rlt2b}  \displaystyle \sum_{k=1 \atop {k\not=i}}^{N} x'_{ijkn} = x_{ij} \ : \ (i,j,n = 1,..,N) \ ;\  n\not=j 
\end{equation}

\begin{equation}
\label{eq:rlt2c} \ \ \ \ \displaystyle \sum_{q=1 \atop {j\not=n}}^{N} x'_{ijkn} = x_{ij} \ :\ (i,j,k = 1,..,N) \ ; \ i\not=k   
\end{equation}

\begin{equation}
\label{eq:rlt2d} \ \ \ \ \ \displaystyle \sum_{p=1 \atop {p\not=i,k}}^{N} x''_{ijknpq} = x'_{ijkn} \ : \ (i,j,k,n,q = 1,..,N) \ ;\  k\not=i; \ \ distinct \ j,q,n \ 
\end{equation}

\begin{equation}
\label{eq:rlt2e} \ \ \ \ \ \displaystyle \sum_{q=1 \atop {q\not=j,n}}^{N} x''_{ijknpq} = x'_{ijkn} \ :\ (i,j,k,n,p = 1,..,N) \ ;\ distinct \ i,k,p \ ; \ n\not=j  
\end{equation}

\begin{equation}
\label{eq:rlt2f} \ \ \ \  x'_{ijkn}=x'_{knij} \  (2\ complementary \ coefficients)\ : \ (i,j,k,n = 1,..,N) \ ; \  i<k \ ; \ j\not=n    
\end{equation}

\begin{equation}
\label{eq:rlt2g} \ \ \ \  x''_{ijknpq}=x''_{ijpqkn}=x''_{knijpq}=x''_{knpqij}=x''_{pqijkn}=x''_{pqknij} \ (6 \ complementary \ coefficients)  : \atop  \ (i,j,k,n,p,q = 1,..,N) \ ; \  i<k<p \ ; \ distinct  \ j,n,q \ 
\end{equation}

\begin{equation}
\label{eq:rlt2h} \ \ \ \ x_{ij} \ge 0\ :\ (i,j = 1,..,N) \ ; \ 
\end{equation}

\begin{equation}
\label{eq:rlt2i} \ \ \ \ x'_{ijkn} \ge 0\ :\ (i,j,k,n = 1,..,N) \ ; \  i<k \ ; \ j\not=,n
\end{equation}

\begin{equation}
\label{eq:rlt2j} \ \ \ \ x''_{ijknpq} \ge 0\ :\ (i,j,k,n,p,q = 1,..,N) \ ; \  i<k<p \ ; \ distinct j,n,q  
\end{equation}

\noindent
and equations (\ref{eq:qap2}) and (\ref{eq:qap3});

To obtain the formulation (\ref{eq:rlt2a} - \ref{eq:rlt2j}), we first apply the steps of RLT1: It's applied  $(N \times (N-1) )^2$ new constraints (\ref{eq:rlt2b}) and (\ref{eq:rlt2c}) by multiplying the constraints  (\ref{eq:qap2}) and (\ref{eq:qap3}) for each of $N^2$ binary variables $x_{kn}$, , with $i \neq k$ and with $j \neq n$, (and changing the indexes  $i$ and $k$ and the indexes $j$ and $n$). After, we replaced each product  $x_{ij}x_{kn}$ by binary variable  $x'_{ijkn}$ and each $x_{ij}x_{ij}$  for  $x_{ij}$.
Because the commutative property of the product, the additional constraints (\ref{eq:rlt2f}) are imposed. The  $x'_{ijkn}$ and $x'_{knij}$ are called of complementary coefficients.

Next, we applied the steps of RLT2: It's added  $(N \times (N-1) \times (N-2))^2$ news constraints  (\ref{eq:rlt2d}) and (\ref{eq:rlt2e}) obtained by multiplying the constraints  (\ref{eq:rlt2b}) and (\ref{eq:rlt2c}) by each of $N^2$ binary variables $x_{pq}$, with distinct $i,k,p$ and also distinct $j,n,q$ , (changing the indexes $i$, $k$ and $p$, and the indexes $j$, $n$ and $q$).  Then, each product $x'_{ijkn}x_{pq}$ is replaced for each binary variable $x''_{ijknpq}$. Similarly the reformulation of RLT1, the constraint (\ref{eq:rlt2g}) also should be imposed. 
The coefficients $x''_{ijknpq}, x''_{ijpqkn}, x''_{knijpq}, x''_{knpqij},  x''_{pqijkn}$, and $x''_{pqknij}$ are called complementary coefficients.

We represent the current dual solution by a set of matrices containing the modified coefficients costs (reduced costs) that are maintained non-negative during the execution of algorithm.
The cost coefficients $b_{ij} \ \forall \ (i,j = 1,..,N)$ are stored in a $N \times N$ matrix  $B$, the cost coefficients $c_{ijkn} \ \forall \ (i,j,k,n = 1,..,N)$, with $i \not=k$ and $j \not=n$ are stored in a  $N(N-1)  \times N(N-1)$ matrix $C$, and finally,  the cost coefficients $d_{ijknpq} \ \forall \ (i,j,k,n,p,q = 1,..,N)$, with distinct $i,k,p$ and also distinct $j,n,q$, are stored in a $N(N-1)(N-2) \times N(N-1)(N-2)$ matrix $D$.

\section{The dual algorithm  to calculate the lower bound} \label{sec:dualascent}

The   Lower Bound, $LB$, is the resulting value from maximizing the objective function obtained from the dualization of the RLT2 formulation applied to the  QAP (Equations (\ref{eq:qap2}), (\ref{eq:qap3}) and (\ref{eq:rlt2a} - \ref{eq:rlt2j})).
The dual algorithm implemented in this work consists of to modify the $LB$ and the coefficients of $B$, $C$ and  $D$ so that no value becomes negative and the cost of some viable solution to the QAP remains unchanged after modification according to the equations (\ref{eq:qap2}), (\ref{eq:qap3}) and (\ref{eq:rlt2a} - \ref{eq:rlt2j}). As result of this property, the $LB$, at any time of algorithm execution, is a valid bound for the cost of the optimal solution.

Our approach to maximize  $LB$ is to transfer costs from $D$ to $C$, from $C$ to $B$, and then, from $B$ to $LB$. For this, we use a modified version of the dual RLT2 algorithm of \cite{adams2007}.  The Algorithm 1 consists of a loop with 3 operations: Costs Spreading, Transfer between Complementary Costs and Cost Concentration.

\begin{algorithm}[htbp]
\Begin{
\SetKwFor{Enqto}{\textit{loop}}{}{End}
  \textit{$LB \leftarrow 0$ }\\
	\textit{$UB \leftarrow $ best known solution researched by a heuristic }\\
	\textit{$K \leftarrow $ minimal progress limit of $LB  ( 0,01\% of UB )$}\\
  \textit{$b_{ij} \leftarrow $ (assignment cost of $i$ at $j \ \forall \ (i,j)$}\\
  \textit{$c_{ijkn} \leftarrow$ $f_{ik}  \times d_{jn} \ \forall \  (i,j,k,n)$ with $i\not=k$ and $j\not=n$}\\
  \textit{$d_{ijknpq} \leftarrow 0$ $\ \forall \ (i,j,k,n,p,q)$ with distinct $ i,k,p $ and also distinct  $j,n,q $}\\
	\textit{$progress \leftarrow 1 $}\\
	\Enqto{  \textit{While ($progress >= K$) and ($LB < UB$) }   }  
  {
    \textit{Costs Spreading from $B$ to $C$} \\
    \textit{Costs Spreading from $C$ to $D$}\\
    \textit{Transfer between Complementary Costs of $D$}\\
    \textit{Cost Concentration from $D$ to $C$ ($c_{ijkn}$  $\leftarrow  Concentrate(d_{ijkn})$)}\\
    \textit{Transfer between Complementary Costs of $C$}\\
    \textit{Cost Concentration from $C$ to $B$ ($b_{ij}$  $\leftarrow  Concentrate(c_{ij})$)}\\
    \textit{Cost Concentration from $B$ to $LB'$ ($LB'$  $\leftarrow  Concentrate( B )$)}\\
    \textit{LB $\leftarrow$ LB + LB'}\\
		\textit{$progress \leftarrow (LB' / UB)$}\\
  }
}
\caption{\textit{  }}
\label{alg:alg7}
\end{algorithm}

The Parameter $K$ is provided at the beginning of the execution of the application and serves to stop the dual loop. The $K$ is the minimum percentage of the $LB$ progress. The $K$  doesn't have a defined value, it depends of the instance, dimension and the history of experiments. The $K$ value is between 0,01\% $(K = 0,00001)$ and $100\% (K = 1)$.

\subsection{Cost Concentration} \label{sec:dual}

The cost concentration operation between the matrices  consists of cost transfer from $D$  to $C$, from $C$ to $B$  and from $B$ matrix to $LB$, obeying the constraints 
 (\ref{eq:qap2} - \ref{eq:qap3}) and   (\ref{eq:rlt2b} - \ref{eq:rlt2e}).  The cost concentration operation is solved as a linear assignment problem. To solve each linear assignment  \cite{adams2007} adopts the hungarian algorithm, \cite{munkres1957}. The hungarian algorithm  minimize the total cost  $S$ of a designation. We consider an example adopting a $M$ cost matrix with dimension  $N \times N$, where each coefficient $M_{rs}$ corresponds to the cost of assigning a facility $r$ to a location  $s$. The objective function is then:  $min\ S = \sum_{r}^{N} \sum_{s}^{N} M_{rs}x_{rs}$ 
where $x_{rs} \in \{0,1\}$,  $\sum_{r=1}^{N} x_{rs} = 1\ \forall \ s\in \{1,..,N\}$ ,  $\sum_{s=1}^{N} x_{rs} = 1\ \forall \ r\in \{1,..,N\}$.

For cost concentration operation from $D$ to $C$, the  $M$ matrix has a dimension $(N-2) \times (N-2)$. For each $(i,j,k,n)$, with $i\not= k$ and $j \not= n$, $M$ receive the $(N-2)^2$ costs elements of  $D_{ijkn}$ submatrix. 
For each  $(r,s = 1,..,N-2)$, $M_{rs}$ receives $d_{(ijkn)pq}$  $(p,q = 1,..,N)$, with $p\not= \{i,k \}$ and $q\not= \{j,n \}$. 
$S$ is obtained by the application of  hungarian algorithm at $M$ and adding $S$ in $c_{ijkn}$. Finally, for each $(i,j,k,n)$, with $i\not= k$ and $j \not= n$, each cost coefficient  $d_{(ijkn)pq}$  is replaced by its correspondent residual coefficient of $M$. We represent this cost transfer operation as: $c_{ijkn}$ $\leftarrow Concentrate(D_{ijkn})$. Similarly, it's  done for cost concentration operation from  $C$ to $B$, represented by $b_{ij}$  $\leftarrow  Concentrate(C_{ij})$, with a $(N-1)^2$ $M$ matrix and the cost concentration operation from  $B$ to $LB$, with a  $N^2$   $M$ matrix, represented this operation by  $LB$ $\leftarrow  Concentrate( B )$.

\subsection{Spreading costs} \label{sec:costs_spreading}

The cost spreading operation is the inverse of cost concentration operation, and is done from  $B$ to $C$ and from $C$ to $D$. The cost spreading operation from $B$ to $C$ consists of:  For each $(i,j)$, the cost element $b_{ij}$ is spreading by the  $(N - 1)$ lines of $C_{ij}$ submatrix, ie each cost element $c_{ijkn}$ receives an increase of $b_{ij} / (N-1)$, $\forall \ (i,j,k,n)$, with $ k\not=i\ $ and $ \ n\not=j$. After the update of $C$, $b_{ij} = 0 \ \forall \ (i,j)$.

Similarly the previous operation, the cost spreading from $C$ to $D$ consists in: for each $(i,j,k,n)$, the cost element $c_{ijkn}$ is spreading by the $(N - 2)$ lines of $D_{ijkn}$ submatrix, ie each cost element $d_{ijknpq}$ receives an increase of $c_{ijkn} / (N-2)$, $\forall \ (i,j,k,n,p,q)$, with distinct $i,k,p$ and also distinct $j,n,q$. After the update of $D$, $c_{ijkn} = 0$ $\forall (i,j,k,n)$, with $\ k\not=i$ and $n\not=j$.

\subsection{Transfer costs between complementary coefficients}  \label{sec:complementary_coefficients}

The complementary coefficients enable communication costs between submatrices. This operation is essential to achieving a good $LB$ by applying the RLT method. A good choice of cost transfer strategy allows to reach tighter $LB$ with low runtime of dual algorithms. 
The charge transfer operation between complementary consists of increment one or more coefficients and decrement the other(s), maintaining the total of decrements and increments, between complementary coefficients, always equal to 0.

\section{Considerations about the GPU} \label{sec:GPU}

The GPU works as a co-processor for the CPU executing massively mathematical calculations. The GPUs are composed of several multiprocessor (or SM) type Single Instruction Multiple Data (SIMD). Each SM has a group of processing units (or SP). The NVidia C2070, for example, has 14 SMs each with 32 SPs.

The programming model Compute Unified Device Architecture (CUDA) enables the development of programs directed to exploit the potential of GPUs. 
The jobs are submitted by the CPU (host) for the GPU (device) through calls with signaling defined by the programming model. These tasks are called  kernel. Each CUDA kernel is executed by a thread. These threads are grouped into blocks, and these in grids. When a CUDA program calls a grid to execute on the GPU, each  blocks of  this grid is directed to an available SM. Upon receiving a block, the SM divides the block in sets of 32 consecutive threads (warps). Each warp executes a single instruction at a time. When a block is finished, another is assigned to the SM.

The GPU memory is organized in a hierarchical way comprising of 4 distinct levels: the \textit{Local} memory present in every SP, the \textit{Shared} memory that can be accessed by any SP of the same SM, \textit{Global} memory that is visible by any SP of all SMs, and finally, the \textit{Constant} that only is accessible for reading by all SPs. Such levels have response times about  2, 2, 600 and 600 clock cycles, respectively.

\section{Branch-and-bound with the dual RLT2 adapted for execution on GPU} \label{sec:BB}

The application containing the branch-and-bound algorithm is executed in the CPU (Host), creating a amount of cpu\_threads (we use this denomination to  trivial threads in order to differentiate of GPU threads) equal the total of available GPUs. The root node of the branch-and-bound tree is performed by cpu\_thread with ID = 0. Next, at the first Branch, each cpu\_thread takes a subtree (or node) and execute a depth-first search. When it finished your subtree, the cpu\_thread takes another node that has not been fathomed.

The  $B$, $C$ and $D$ matrices are stored in GPU Global memory. For each fathomed node the cpu\_thread executes the dual algorithm of Section \ref{sec:dualascent}. The cost concentration, costs spreading and costs transfer between complementary coefficients  operations are executed by GPU.

The cost concentration use a linear assignment algorithm which could be the Hungarian algorithm, \cite{munkres1957}, widely used due to its efficiency, but this algorithm doesn't allow a good parallelization.
We choose to use the Auction Algorithm  \cite{leilao1989}, also used as a linear assignment algorithm and is more efficient for parallelization on GPU. Each person (auction participant) and each object of Auction algorithm corresponds to, here in this work, the facility  and the location of QAP original problem, respectively. The parallelization strategy is to let each GPU thread with the function of a person on Auction Algorithm.

For the cost concentration operation from $D$ to $C$, a $(N-2) \times (N-2)$ $M$ matrix is allocated in Shared memory of each SM. 
For each $(i,j,k,n)$,  the cost coefficients of submatrix $D_{ijkn}$ are transferred to $M$ and a auction algorithm is executed. 
It's allocated one warp (32 threads) for each auction algorithm execution, and at the end, the residual coefficients of  $M$ matrix are transferred, from the Shared memory  to  $D_{ijkn}$ on Global memory. This procedure is same for the cost concentration operation from $C$ to $B$ and from $B$ to $LB$.

We opted not to implement the sharing of complementary coefficients at the same memory location, possible by the constraints (\ref{eq:rlt2f}) and  (\ref{eq:rlt2g}). This resourse was used in \cite{adams2007}, and could reduce the required memory to store the $D$ matrix. 
 The reason of our option is the loss of application performance, since the coefficients of the matrices would not be in contiguous blocks, that would spend several read cycles to transfer from Global memory to $M$ matrix in Shared memory. Another reason, which also causes loss of performance, is that  several threads of distinct warps  concurrently accessing the same location in memory, and thus, serializing access to memory.
 
We note that each coefficient of submatrix $(d_{ijkn})_{pq}$ corresponds to complementary coefficient of submatrix $(d_{knij})_{pq}$. 
Then, we introduced the concept of complementary submatrices where: for all $(i,k = i,...,N)$, and $(j,n = 1,...,N)$, with $i<k$, 
the submatrices $(d_{ijkn})_{pq}$ and  $(d_{knij})_{pq}$ are complementaries. Using this concept, we can reduce by a half the required memory to store the $D$ matrix, as also reduce by half the total of costs concentration operations from $D$ to $C$ impacting the application runtime.

The Branch strategy consists of a strong branch that is done a preliminary evaluation of a set of assignments and concentrations of the facilities not yet assigned to each location also unassigned, and then, it is selected a row or column that obtains the highest $LB$. In the strong branch the costs concentration follows the RLT1 dual algorithm similar to Section \ref{sec:dualascent},  except by the operations with transfer costs of $D$ matrix.

\section{Experimental Results}\label{sec:Resultados}

The proposed algorithm, here called GPU dual RLT2,  was implemented using the C++ language and the CUDA programming model. The experiments were executed on a non-exclusive  machine with a Intel Xeon Hexcore CPU with 24 GB of RAM and one NVidia GeForce GTX TITAN GPU. This GPU has 3,072 of  1,000 MHz SPs, organized in 24 SMs (128 SPs each) and 12 GB DDR5 Global memory.

\begin{sidewaystable} 



\centering
\begin{tabular}{l|c|c||c|c|c||c|c|c|c||c|c|c|c} \hline
         &    &          &  \multicolumn{3}{c||}{dual RLT2} & \multicolumn{4}{c||}{RLT1/2/3 Parallel C} &  \multicolumn{4}{c}{GPU dual RLT2}  \\   \cline{4-14}     
Instance & N  & Optimal    &  Nodes    & Memory& time   & \multicolumn{2}{c|}{Nodes}& Memory& time   &	Nodes   & \multicolumn{2}{c|}{Mem. (GB)} &time \\ \cline{4-14}
         &    &          & B\&B    &  (GB)  &  (s)    &   B\&B  &  RLT3 &  (GB)  &   (s)   &  B\&B    &  \textit{ Host} & \textit{Device} & (s)    \\    \hline \hline
nug20    & 20 & 2570     & 1,407   &   0.4  & 2,978   &  39    & 9      & 18.4   &  736    & 565 & 1      &  0.4   & 6	   \\ 
nug22    & 22 & 3596     & 1,450   &   0.8  & 3,361   &  52    & 16     & 40.7   &  1,578  & 927 & 1      &  0.6   & 11		 \\
nug24    & 24 & 3488     & 37,099  &   0.8  & 5,781   &  102   & 16     & 85.0   &  5,097  & 606 & 1,5    &  1.2   & 29	 \\
nug25    & 25 & 3744     & 15,497  &   1.7  & 124,702 &  267   & 31     & 120.3  & 23,072  & 3,616   & 2      &  1.5     & 491 	 \\ 
nug27    & 27 & 5234     &    -    &  -     & -       &  359   & 46     & 231.5  & 53,227  & 1,831   & 3      &  2  & 360 \\ 
nug28    & 28 & 5166     & 202,295 &  3.6   &2,856,392&  1,538 & 115    & 321.4  & 130,527 & 7,269   & 4      &  3    & 2,084 \\ 
nug30    & 30 & 6124     & 543,061 & 6      &19,735,563&  3,383& 251    & 722.8  & 733,812 & 103,380    & 5      &  6    & 88.867 \\ \hline
tai20a   & 20 & 703482   &  -      &   -    &         & -      &   -    & -      &    -    & 7,117   & 1      &  0.4   & 71 \\ 
tai20b   & 20 & 122455319&  -      &   -    &         & -      &   -    & -      & -       & 211     & 1      &  0.4   & 6	 \\
tai25a   & 25 & 1167256  &  -      &   -    & -       &  -     &   -    & -      &  -      & 324,394    & 2      &  1.7    	& 109,011 \\ 
tai25b   & 25 & 344355646&  -      &   -    & -       &  -     &   -    & -      &  -      & 350   & 2      &  1.7    	& 63 \\ 
tai30b   & 30 & 637117113&  -      &   -    & -       & W/O inf& W/O inf& 753    & 315.584 & 1,011  & 5      &  6    & 448 \\ 
\textbf{tai35b}   & \textbf{35} & \textbf{283315445*}&  -      &   -    & -      &        &        &        &         & \textbf{1,029,312}   & \textbf{5}  &\textbf{9}    & \textbf{473,873} \\ 
\textbf{tai40b}   & \textbf{40} & \textbf{637250948*}&  -      &   -    & -      &        &        &        &         & \textbf{2,510,362
}   & \textbf{5}  &\textbf{12}    & \textbf{4,949,444
} \\ 
\hline
kra30a   & 30 & 88900    &  -      &   -    & -       &  -     &   -    & -      &   -     & 20,764  & 4      &  6   	& 940 \\ 
kra30b   & 30 & 91420    &  -      &   -    & -       &  -     &   -    & -      &   -     & 24,082  & 4      &  6   	& 16,552 \\ \hline
tho30    & 30 & 149936   &  -      &   -    & -       &  -     &   -    & -      &   -     & 22,429  & 4      &  6   	& 17,219 \\ \hline
\multicolumn{14}{l}{\textbf{* Instances  solved exactly for the first time}}
\end{tabular}
\label{Tabela1}
\caption{ Comparison between the results of \cite{hahn2012} with the  dual RLT2 of \cite{adams2007}, the results of  RLT 1/2/3 Parallel C algorithm of \cite{hahn2013} and  the results of GPU dual RLT2 algorithm proposed in this work}
\end{sidewaystable}

Table 1 shows the results achieved for different QAP instances. We used 2 others works as reference to evaluate our experiments,  the sequential dual RLT2 algorithm of \cite{adams2007} and the RLT1/2/3 Parallel C algorithm  of \cite{hahn2013}. Recent experiments were done with the dual RLT2 algorithm and the new results were published in  \cite{hahn2012} and we put them in Table 1. The dual RLT2 algorithm experiments of \cite{hahn2013} were made in two environments:  one machine with a 1.9GHz E6900 CPU, used to process the Nug20, Nug22, Nug24 and Nug25 instances, and one machine with a  733MHz Itanium  CPU  used to process the Nug28 and Nug30 instances. The RLT1/2/3 Parallel C algorithm experiments of \cite{hahn2013} were executed in  30 of the 64 hosts of Palmetto Supercomputing Cluster at the University of Clemson, with 2TB of shared memory and each host with one 2.2GHz Intel Xeon CPU.

Our proposal was evaluated with instances found in QAPLIB, \cite{qaplib2}. Besides the instances used in \cite{hahn2013} we added others as: Tai20a, Tai20b, Tai25a, Tai25b, Tai30b, Tai35b, Tai40b, Kra30a, Kra30b and Tho30.

Table 1 is organized as follows: the initial three columns has the information about the instances as identification, the total of facilities (N) and the optimal value found in the literature. The other columns report the results of the algorithms cited above. In the fourth, fifth and sixth column, the amount of nodes of branch-and-bound, the total of main memory and the runtime, respectively, for the dual RLT2 algorithm. In the seventh, eighth, ninth and tenth column, the number of nodes of branch-and-bound, the total of nodes in RLT3 step, the amount of main memory and the runtime, respectively. Finally, the results obtained in our algorithm (GPU dual RLT2), in sequence: the total of fathomed nodes, the amount of RAM used in the \textit {host}, the total of Global memory used in GPU, and the runtime clock.

Comparing the results achieved by dual RLT2  and the GPU dual RLT2 we observed that the speedup is between 50 to 300 depending of the instance. Some instances have the B\&B subtrees very unbalanced that some subtrees are finished earlier than others, then, load balancing procedures were implemented.

The RLT1/2/3 Parallel C algorithm extends the relaxation to the level 3 RLT reaching tighter bounds that dual RLT2, however, there is the need of an extensive working memory. Besides the $B, C$ and $D$ matrices of RLT2, the implementation of RLT3 requires a $(N \times (N -1) \times (N-2) \times (N-3))^2$ $E$ matrix that to require a large amount of working memory for storing and processing the operations with this matrix .

Comparing the results of the GPU dual RLT2 algorithm with the achieved by the RLT1/2/3 Parallel C algorithm, we observed that the runtime is about 140x lower for instances as Nug20, Nug22, Nug24  and Nug27, approximately 50x lower for Nug25 and Nug28 and about of 8.5x lower for Nug30. 
The reduction in performance on experiments of RLT2 in relation to RLT3 with increasing size of the problem is pointed out in \cite{HahnRLT3}. They show that the runtime of one instance larger than 28 facilities in applications based in RLT2 relaxation are superior than applications with relaxation RLT3.

\begin{table}[!htb] 
\scriptsize
\centering
\begin{tabular}{l|c|c} \hline
Instance &  Optimal  &  Solution  \\    \hline \hline
\textbf{tai35b}  & \textbf{283315445}&      14  12   5  18  10  30  11  22   1  19   9  20  32  17   2  33   3   8  13  27  16   4  34   7  23  24   6  35  31  28  15  21  29  26  25\\ \hline
\multirow{2}{*}{\textbf{tai40b}}  & \multirow{2}{*}{\textbf{637250948}}&   
36  1 15 11 25 37 31 19 39 13 27  7 40 22  4 33 16 34 10 14\\  
& & 12 23  5 32 35 38  9  3 30 29 24 17  2  6 28  8 20 26 18 21\\ 
\hline
\end{tabular}
\label{Tabela2}
\caption{ Instances solved exactly by first time}
\end{table}

In Table 1, the Instances Tai35b and Tai40b  in bold were resolved exactly by first time  confirming the optimal value obtained  by heuristics. The solutions reached are present in Table 2.  To solve these instances the GPU dual RLT2 algorithm was executed on a machine without exclusivity and subject to interruptions due to lack of energy and decreased network, imposing the need to implement a checkpoints procedure. The runtime shown in Table 1 corresponds to the sum of the individual clock times of execution and each re-execution after the interrupts. The gap reached in  Branch-and-bound root node also were the lower of literature, they were 4.92 \% and 4.14 \% for tai35b and tai40b, respectively,  the best known found in QAPLIB site were 14.52 \% and  11.43 \%.

\section{Conclusions and future works} \label{sec:conclusao}

The recent commercial GPUs have the tendence a high increasing of amount of SPs, however, their global memory have not the same growth,  this scenario provides advantages to applications with relaxation RLT2 over the applications with RLT3.


The  GPU dual RLT2 algorithm is still under review, and new changes now allow better performance. A distributed version is being evaluated promising good results.  Results and new contributions will be presented in future work.

\bibliographystyle{sbpo}
\bibliography{ref1}

\end{document}